\documentclass{PoSLike}

\title{Progress Report on an Ultra-compact LumiCal}

\ShortTitle{Progress Report on an Ultra-compact LumiCal}

\author{\speaker{Oleksandr Borysov} \\
        Tel Aviv University (IL)\\
        E-mail: \email{oleksandr.borysov@cern.ch}}

\author{On behalf of the FCAL collaboration}

\abstract{A new design of a detector module of submillimeter thickness for an electromagnetic calorimeter is presented. It is aimed to be used in the luminometers LumiCal and BeamCal in future linear e$^{+}$e$^{-}$ collider experiments. The module prototypes were produced utilizing novel connectivity schemes technologies. They are installed in a compact prototype of the calorimeter and tested at DESY with an electron beam of 1~GeV~--~6~GeV. The performance of eight detector modules and the possibility of electron and photon identification is studied.}

\FullConference{Talk presented at the International Workshop on Future Linear Colliders (LCWS2016),\\
                  Morioka, Japan,\\
                  5-9 December 2016.\\
                  C16-12-05.4.}

\begin{document}


\section{Introduction}
\label{Introduction}
Forward calorimeters for future electron positron linear collider experiments have challenging requirements on geometrical compactness and high precision measurements of luminosity~\cite{FCAL_ILC} resulting in the design of highly compact calorimeters. Two such calorimeters are considered to be installed in the forward region of both ILC~\cite{ILC_TDR_v4_det, ILC_TDR_v1_phys} detectors ILD and SiD, and also in the CLIC detector~\cite{CLIC_CDR, CLIC_UPDATE_YP}. Precise measurement of integrated luminosity is provided by the LumiCal detector. Another detector, BeamCal, is designed for instant luminosity measurement and assisting beam tuning when included in a fast feedback system. Both detectors extend the capabilities of the experiments for physics study in the high rapidity region. 

\begin{figure}[h!]
  \begin{minipage}[c]{0.50\textwidth}
    \includegraphics[width=\textwidth]{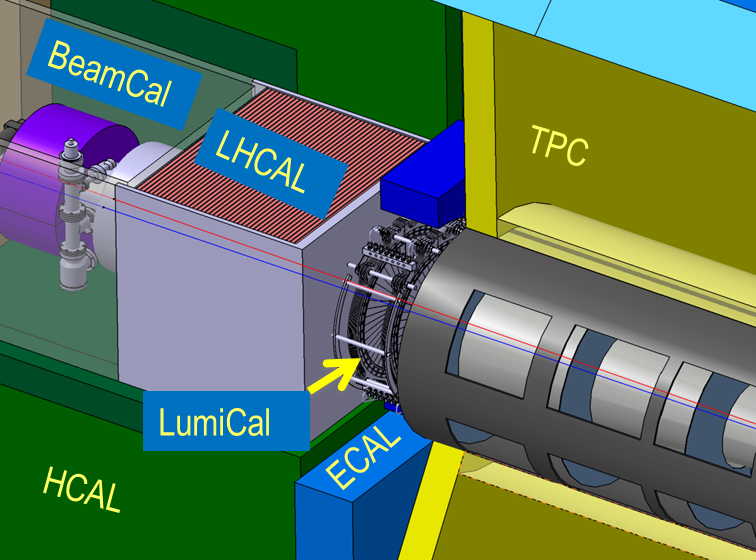}
  \end{minipage}\hfill
  \begin{minipage}[t]{0.45\textwidth}
    \caption{The very forward region of the ILD detector. LumiCal, BeamCal and LHCAL are carried by the support tube 
             for the final focusing quadrupole QD0 and the beam-pipe. TPC, ECAL and HCAL are barrel detectors.
    } \label{fig_forward_ild}
  \end{minipage}
\end{figure}

The layout of one arm of the forward region of the ILD detector is presented in~Fig.~\ref{fig_forward_ild}. LumiCal is positioned in a circular hole of the end-cap electromagnetic calorimeter ECAL. BeamCal is placed just in front of the final focus quadrupole. LumiCal is designed as a sampling calorimeter composed of 30 layers of 3.5~mm (1X$_0$) thick tungsten absorbers and silicon sensors placed in a one millimeter gap between absorber plates. BeamCal has a similar design as LumiCal, but the sensors must withstand much higher radiation doses. For the current BeamCal baseline design, GaAs sensors are considered. The similarity between LumiCal and BeamCal designs implies that the technology developed for one of them can be effectively used for the other. The present report describes the progress in construction and test of detector modules for LumiCal built using silicon pad sensors. 

\begin{figure}[h!]
  \begin{minipage}[t]{0.47\textwidth}
    \includegraphics[width=\textwidth]{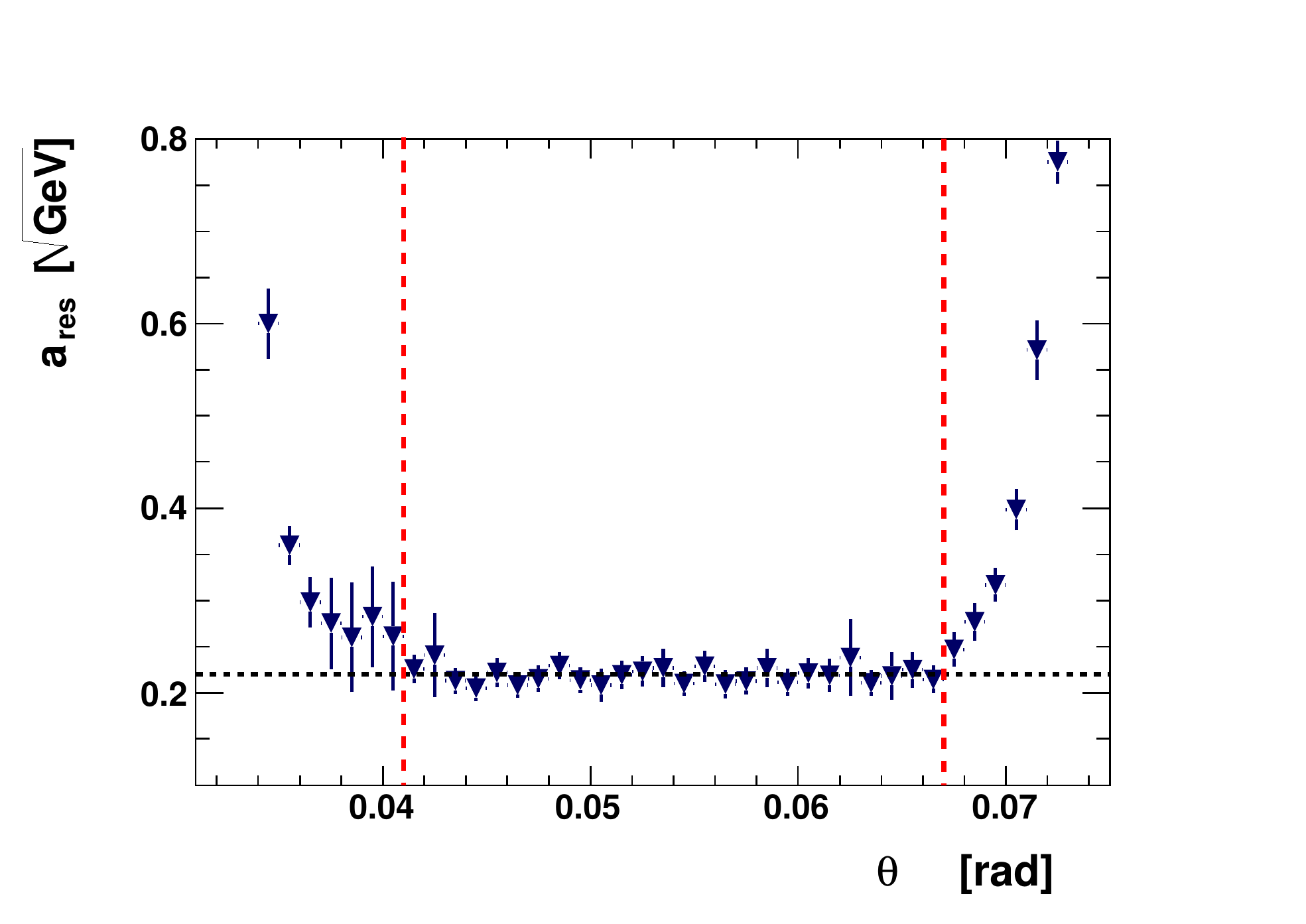}
    \caption{The energy resolution, a$_{\rm{res}}$, for 250~GeV electrons as a function of the polar angle, $\theta$, covering the polar angle range of the LumiCal~\cite{FCAL_ILC}.}
    \label{fig_fidvol_theta}
  \end{minipage}\hfill
  \hspace{0.06\textwidth}
  \begin{minipage}[t]{0.43\textwidth}
    \includegraphics[width=\textwidth]{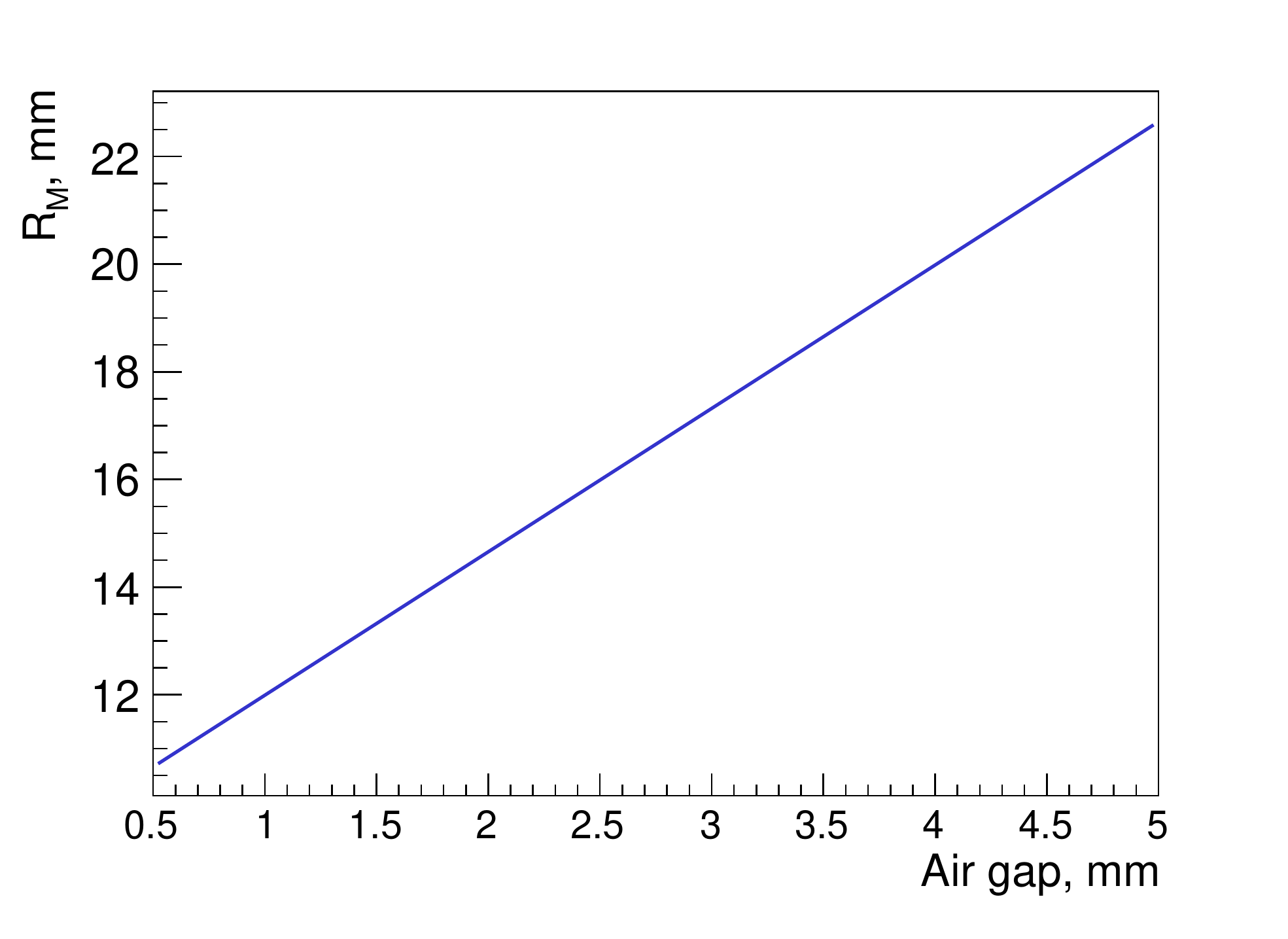}
    \caption{The Moli\`ere radius of a stack of 3.5~mm thick~($1X_0$) tungsten absorber plates as a function of the air gap between them.}
    \label{fig_MR_tungsten_air}
  \end{minipage}\hfill
\end{figure}

Luminosity in LumiCal is measured using Bhabha scattering, e$^{+}$e$^{-}\rightarrow$~e$^{+}$e$^{-}$($\gamma$), as a gauge process. Its cross section can be precisely calculated in QED and the luminosity $\sf{L}$, is obtained as 
\begin{equation}{
  {\sf{L}} = \frac{{N}_{\mathrm{B}}}{\sigma_{\mathrm{B}}},
  }\label{lumiDefEQ} 
\end{equation}
where $N_{\mathrm{B}}$ is a number of Bhabha events registered by LumiCal in a given range of polar angles~($\theta$) and $\sigma_{\mathrm{B}}$ is the integral of the differential cross section over the same range ($\theta_{min}$,~$\theta_{max}$). This range is defined by the fiducial volume of the calorimeter and for the LumiCal baseline design it was determined in a simulation~\cite{FCAL_ILC}. Fig.~\ref{fig_fidvol_theta} shows the energy resolution of LumiCal as a function of the polar angle of the scattered electron with vertical lines indicating $\theta_{min}$ and~$\theta_{max}$ of the fiducial volume for the polar angle. The value~${{a}}_{\rm{res}}$ shown on the plot is a parameter which is typically used to parameterize the relative energy resolution of an electromagnetic calorimeter, 
\begin{equation}{
  \frac{\sigma_{{E}}}{{E}} = \frac{{{a}}_{\rm{res}}}{\sqrt{{{E}}_{\rm{beam}}~\mathrm{(GeV)}}},
  }\label{engyResEQ} 
\end{equation}
where $E$ and $\sigma_{E}$ are, respectively, the central value and the standard deviation of the distribution of the energy deposited in the calorimeter for a beam of electrons with energy ${E}_{\rm{beam}}$. Points outside the fiducial range illustrate the energy lateral leak which is defined by the transverse electromagnetic shower development. The transverse size of the shower is characterized by the Moli\`ere radius. It can be estimated using the formula recommended by PDG~\cite{pdg} for composite materials:
\begin{equation} \label{eq_MR_WF}
\frac{1}{R_\mathcal{M}} = \frac{1}{E_s} \sum{\frac{w_j E_{cj}}{X_{0j}}} = \sum{\frac{w_j}{R_{\mathcal{M}j}} } \ ,
\end{equation}
where $j$ runs over elements composing the material or construction, $w_j$ is a mass fraction of the element and $R_{\mathcal{M}j}$ its Moli\`ere radius. 
As an example, Fig.~\ref{fig_MR_tungsten_air} shows the Moli\`ere radius of the structure with tungsten absorber plates of 3.5~mm thickness as a function of the distance between them. This tungsten-air structure is similar to LumiCal and it demonstrates the importance of the compact design of LumiCal with small gaps between absorbers to keep the transverse size of the shower small and to achieve a sufficiently large fiducial volume in polar angles. 

The statistical uncertainty of the luminosity measurement is defined by the number of Bhabha events registered by LumiCal. Taking into account that the differential Bhabha cross section is proportional to $\theta^{-3}$, the number of events in LumiCal is proportional to ($\theta_{min}^{-2}-\theta_{max}^{-2}$). It means that keeping the fiducial volume in polar angle as large as possible allows reducing the statistical uncertainty of luminosity measurement. A small Moli\`ere radius also improves the efficiency to detect electrons on top of a widely spread background produced by beamstrahlung.

In addition, the compact construction of LumiCal and BeamCal are essential to match the strict geometrical constraints imposed by the design of the detectors and accelerator-needs near the interaction point. The present report describes a new design of the LumiCal module which meets the geometrical requirements of LumiCal and its beam test in a compact calorimeter prototype.
%
\section{Thin LumiCal Module Design}
\label{LumiCal_Thin_subsection}
\begin{figure}[h!]
  \begin{minipage}[c]{0.55\textwidth}
    \includegraphics[width=\textwidth]{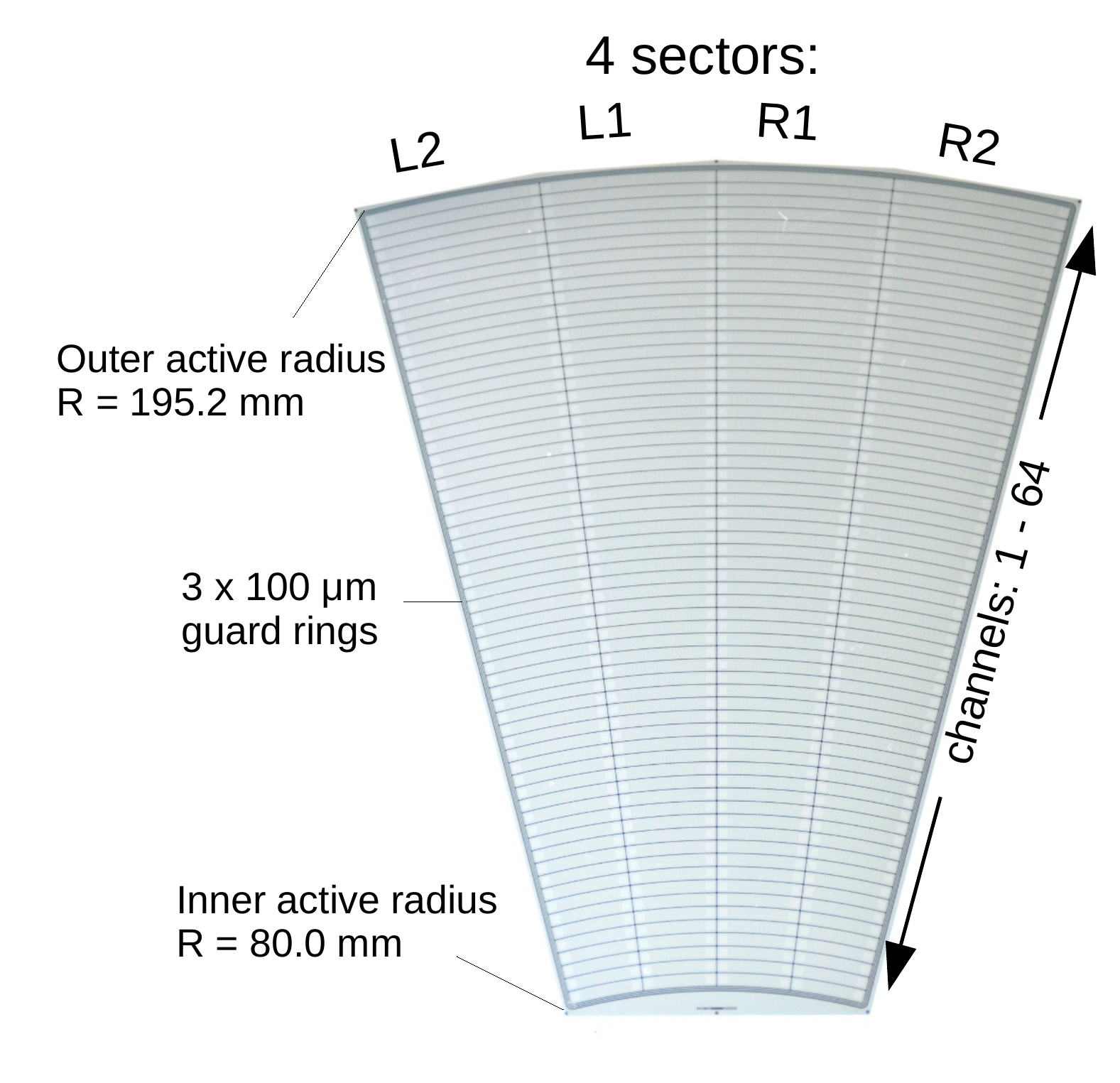}
  \end{minipage}\hfill
  \hspace{0.02\textwidth}
  \begin{minipage}[t]{0.42\textwidth}
    \vspace{30pt}
    \caption{A LumiCal silicon pad sensor.}
    \label{fig_si_sensor}
  \end{minipage}
\end{figure}

The design of a LumiCal sensor was optimized in simulations to provide the required resolution of the polar angle reconstruction. Its picture with structural elements is presented in Fig.~\ref{fig_si_sensor}. The sensor is made of a 320~$\mu$m thick high resistivity n-type silicon wafer. It has the shape of a sector of an angle of 30$^{\circ}$, with inner and outer radii of the sensitive area of 80~mm and 195.2~mm, respectively. It comprises four sectors with 64 p-type pads of 1.8~mm pitch. The sensitive area is surrounded by three guard rings. 

The properties of the sensor were studied in the lab and beam tests. Results of beam tests and more details about the sensor can be found in~\cite{TB2010_jinst, TB2014_Whistler_proc}. The first prototype of the LumiCal detector module, which has been successfully tested in a multilayer configuration, had a thickness of about 4~mm and only 32 pads connected to the readout electronics. 

This work is aimed to use the same silicon sensor for designing and prototyping a detector module of submillimeter thickness suitable for LumiCal. In the module construction, the bias voltage is supplied to the n-side of the sensor, all 256~pads of the sensor are connected to the front-end electronics and the inner guard ring is grounded. In addition, the module should be mechanically solid enough to allow easy handling during lab and beam tests. 

Since the multichannel version of the dedicated front-end electronics is still under development, the APV25 chip~\cite{APV} hybrid board used by the silicon strip detector of the CMS experiment was chosen as a temporary solution. It has 128 channels and two boards read the whole LumiCal sensor.  

\begin{figure}[!h]
  \begin{minipage}{0.62\textwidth}
    \centering
    \includegraphics[width=0.99\columnwidth]{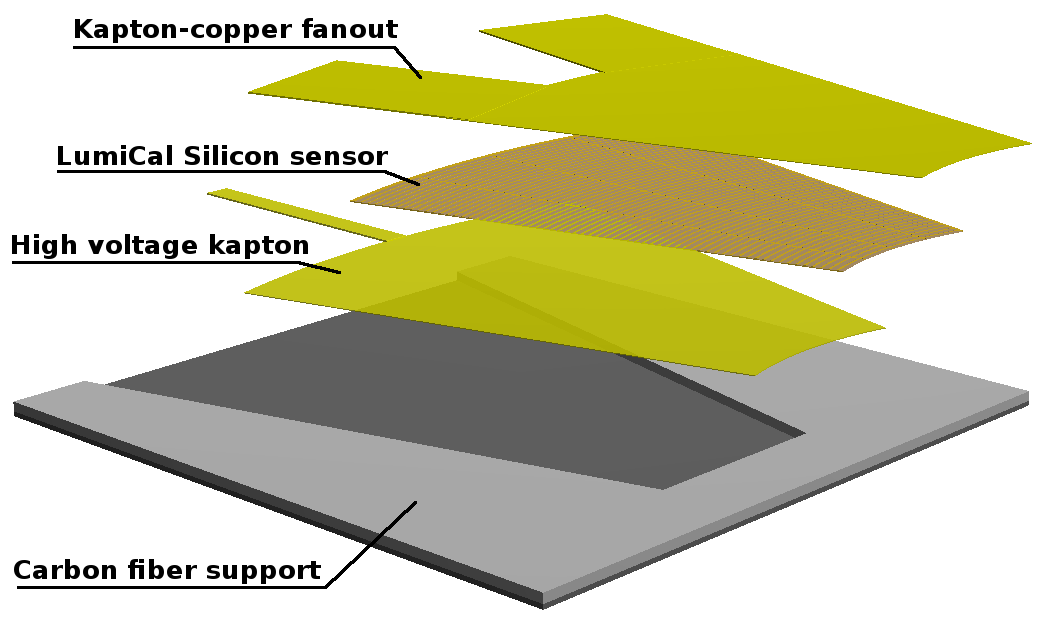}
  \end{minipage}%
  \hspace{0.03\textwidth}
  \begin{minipage}{0.28\textwidth}
      \begin{tabular}{ ll } \hline
              Layer     & Thickness   \\ \hline
              Fan-out    & 120   \\ 
              Adhesive  & 10	  \\ 
              Si sensor & 320 	\\ 
              Adhesive  & 15	  \\ 
              HV Kapton & 70    \\ 
              Adhesive  & 15	  \\ 
              Support   & 100  	\\ \hline
              Total:    & 650	$\mu$m  \\ \hline
      \end{tabular}
  \end{minipage}%
  \caption{Thin LumiCal module assembly. The table shows the thickness of each layer in $\mu$m and the total anticipated thickness of the module.}
  \label{fig_ThinLCAssembly}
\end{figure}

\begin{figure}[h!]
  \begin{minipage}[t]{0.48\textwidth}
    \hspace{0.2\textwidth}
    \begin{minipage}[t]{0.61\textwidth}
      \includegraphics[width=\textwidth]{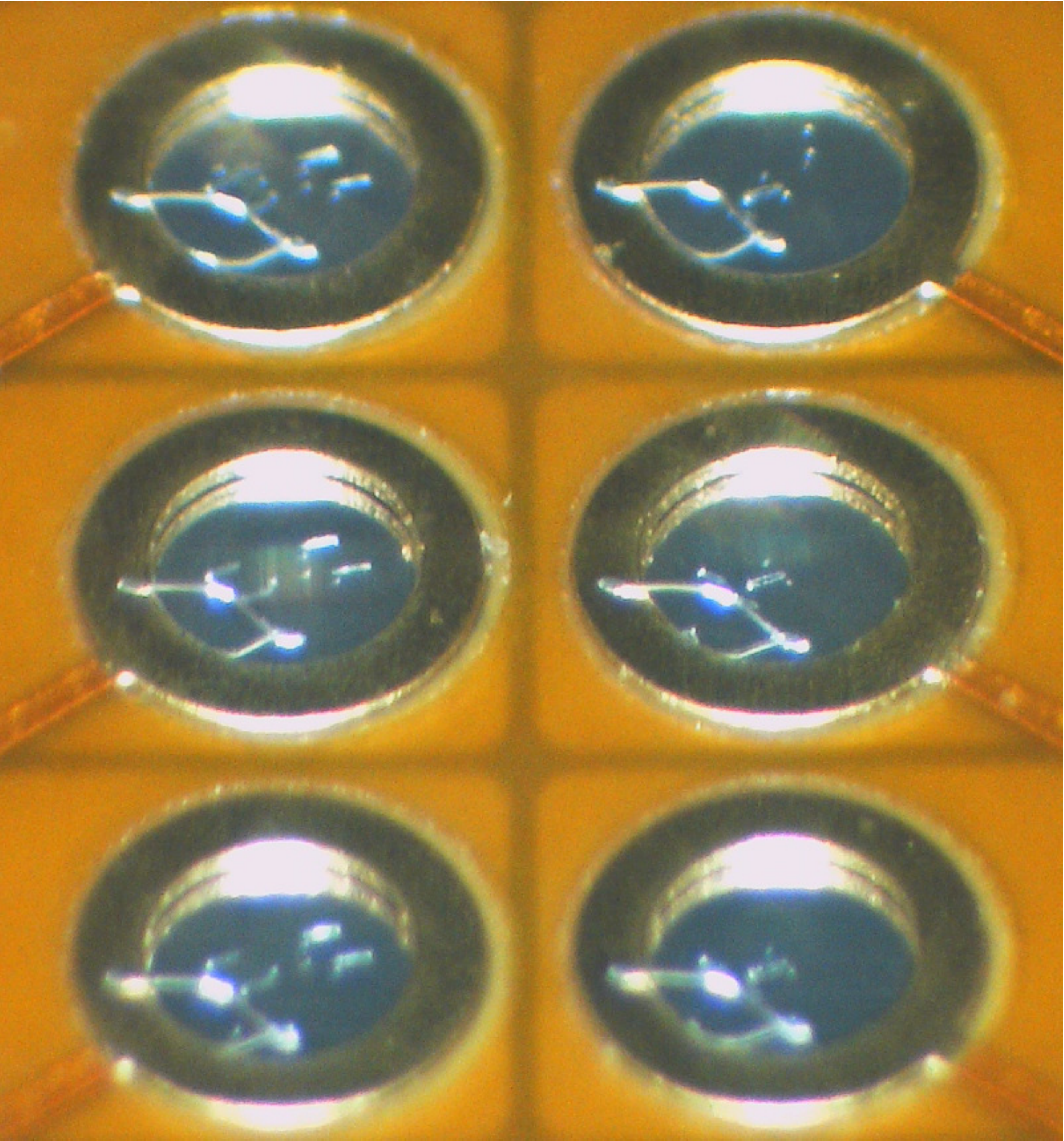}
    \end{minipage}
    \caption{Wire bonding connecting fan-out with LumiCal sensor pads. The middle area of the sensor between sectors L1 and R1 is shown.}
    \label{fig_wire_bonding_lumical}
  \end{minipage}\hfill
  \hspace{0.04\textwidth}
  \begin{minipage}[t]{0.47\textwidth}
    \includegraphics[width=\textwidth]{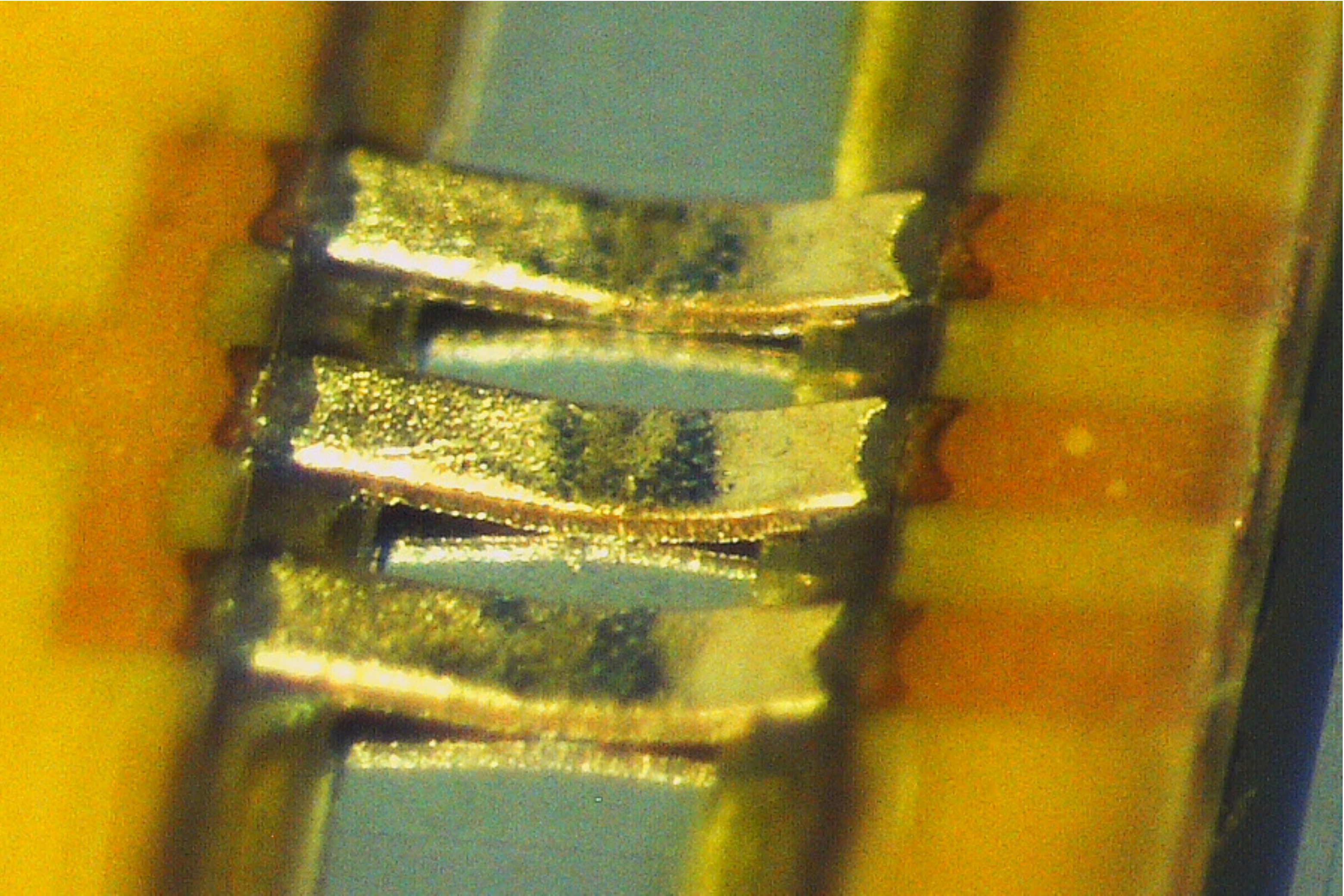}
    \caption{TAB bonding contact between fan-out and a LumiCal sensor pad.}
    \label{fig_tab_lumical}
  \end{minipage}\hfill
\end{figure}

Powering circuits and the fan-out part which connects silicon sensor pads with the front-end board inputs were made from flexible Kapton-copper foils with thickness of 70~$\mu$m for the high voltage one, applied to the back n-side of the sensor and about 120~$\mu$m for the fan-out~(Fig.~\ref{fig_ThinLCAssembly}). The high voltage part was glued to the sensor with a conductive glue, after that the fan-out was glued to the front side of the sensor using epoxy and then ultrasonic wire bonding was used to connect conductive traces on the fan-out to the sensor pads~(Fig.~\ref{fig_wire_bonding_lumical}). Special fixtures were designed and produced to ensure the necessary thickness and uniformity of three glue layers between different components of the LumiCal detector module all over the area of the senor. 

The ultrasonic wire bonding proved to provide good electrical performance, but for a module thinner than 1~mm, the wire loops, which are typically 100$-$200~$\mu$m high, cause a serious problem when the module needs to be installed in a 1~mm gap between absorber plates. The parameters of the bonding machine were studied and tuned to make the loop as low as possible and technically acceptable. The sampling based measurements which have been done using a confocal laser scanning microscope show that the loop height is in the range from 50~$\mu$m to~100~$\mu$m.

Several alternative solutions were considered for connecting sensors with front-end electronics to avoid prone to damage wire loops. One of them is the tape automatic bonding (TAB) technique, which was successfully used for silicon strip detectors for the STAR and the ALICE experiments~\cite{TAB_Star, TAB_Alice}. This technique requires high geometrical accuracy of the bonding parts and it is known that failed bonds are almost not repairable. In order to use TAB for the LumiCal module, the fan-out is redesigned and windowed leads are introduced in the bonding area. The bonding pad of the LumiCal sensor is 1$\times$1~mm$^2$ and it allows several windowed leads to be allocated to cope with possible bonding failures. The TAB fan-out is reduced in size to cover only two sectors of the sensor so that two parts, one of which is flipped with respect to the other, cover all four sectors of the sensor. The same Kapton-copper flexible foil is used for the TAB fan-out. Figure~\ref{fig_tab_lumical} shows three leads of TAB fan-out bonded to the LumiCal sensor. One prototype of the LumiCal module was assembled and prepared for the beam test. The absence of the fragile wire bonding loops on the surface of the module allows to handle it safely without support structure which effectively reduces the module total thickness by another 100~$\mu$m. 

It is worth noting that fan-out and high voltage Kapton parts for the module assembly are produced by industrial companies using their standard processes. With some additional study and optimization, the total module thickness can be further reduced.  

Two approaches are tested for constructing the sensor mechanical support structure. One is made using 3D printing and another is produced from a carbon fiber composite. Two mechanical prototypes of the detector modules were assembled using these two supporting structures and fake sensors. Both approaches allow producing detector module with less than 1~mm thickness, 0.8~mm for 3D and 0.9~mm for the carbon fiber composite. The mechanical stability of the prototype made with carbon fiber composite is significantly better and that method is chosen for the detector production. For the LumiCal module prototype the carbon fiber support was thinned down to 100~$\mu$m in the sensor gluing area, while the surrounding part remained 600~$\mu$m.

\section{Beam Test of the LumiCal Prototype}
\label{TB_setup}
Prototypes of thin LumiCal modules were installed in the 1~mm gap between the tungsten absorbers and tested in two beam test campaigns in 2015 and 2016 at the DESY-II Synchrotron with secondary electrons with energies between 1~GeV and 6~GeV. 

Four modules were assembled for the 2015 beam test which aimed to study the performance of the new design of the modules in a compact calorimeter prototype. For the beam test of 2016, four additional modules were prepared including one assembled using TAB technology. In addition, in 2016 the concept of tracking detectors in front of LumiCal was studied, considered as a tool for electron and photon identification in linear collider experiments. The influence of additional layers of tracking detectors in front of LumiCal was estimated in a simulation~\cite{fcal_25_proc}. It was shown that the tracker does not degrade the shower reconstruction in LumiCal. The efficiency of electron and photon identification can be studied experimentally taking into account back-scattering of particles. 

\begin{figure}[h!]
  \includegraphics[width=\textwidth]{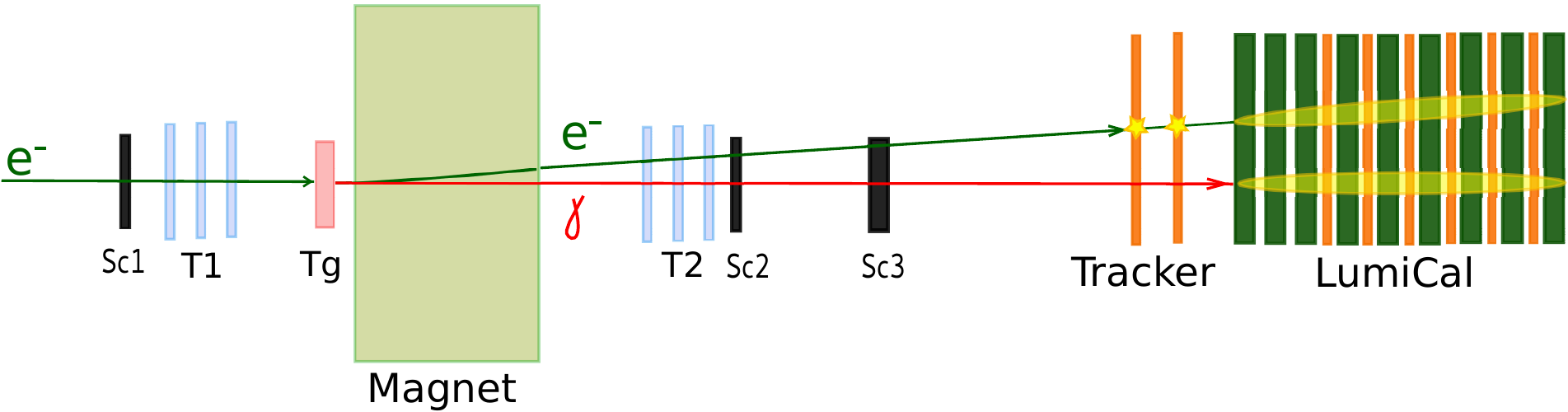}
  \caption{Diagram of the beam test setup and measurements (not in scale). Sc1, Sc2 and Sc3 are scintillator counters; 
           T1, T2~--~three pixel detector planes and Tg~--~the copper target for bremsstrahlung photon production.}
  \label{tb2016}
\end{figure}

\begin{figure}[h!]
  \begin{minipage}[t]{0.45\textwidth}
    \includegraphics[width=\textwidth]{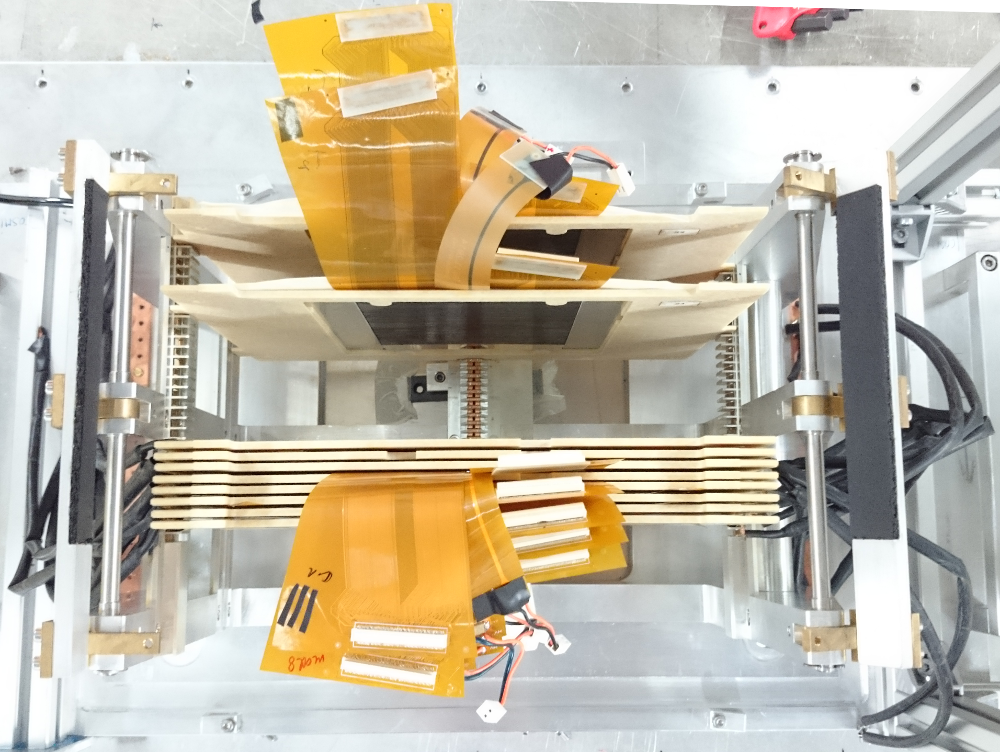}
    \caption{Top view on the assembled LumiCal prototype.}
    \label{tb2016_LumiCal}
  \end{minipage}\hfill
  \hspace{0.03\textwidth}
  \begin{minipage}[t]{0.45\textwidth}
    \includegraphics[width=\textwidth]{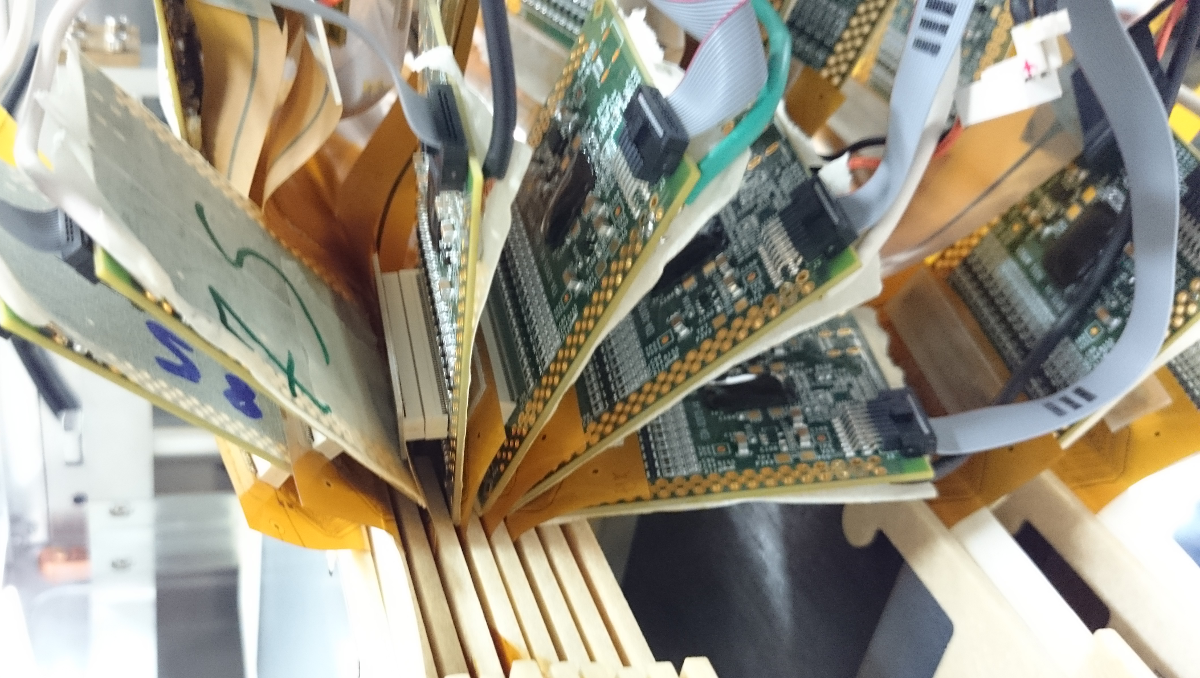}
    \caption{An APV25 hybrid front-end board connected to LumiCal modules.}
    \label{apv_connected_photo}
  \end{minipage}
\end{figure}

The geometry of the setup for this study is optimized in a simulation prior to the beam test and the scheme of the experiment is shown in Fig.~\ref{tb2016}. Bremsstrahlung photons are produced by the electron beam hitting the copper target installed upstream close to the dipole magnet. The magnetic field is chosen to allow both photons and electrons to travel within the acceptance of the second telescope T2 and arrive to LumiCal at a distance between them large enough to be resolved in the calorimeter. For triggering such events with low energy photons, all scintillator counters (Sc1, Sc2 and Sc3) are used in coincidence. In order to register a high energy photon in LumiCal, the last scintillator counter (Sc3) is used in anticoincidence. In this trigger configuration, only events with electrons of low energy and thus large deflection in the magnetic field are recorded.
\begin{figure}[h!]
  \begin{minipage}[t]{0.50\textwidth}
    \includegraphics[width=\textwidth]{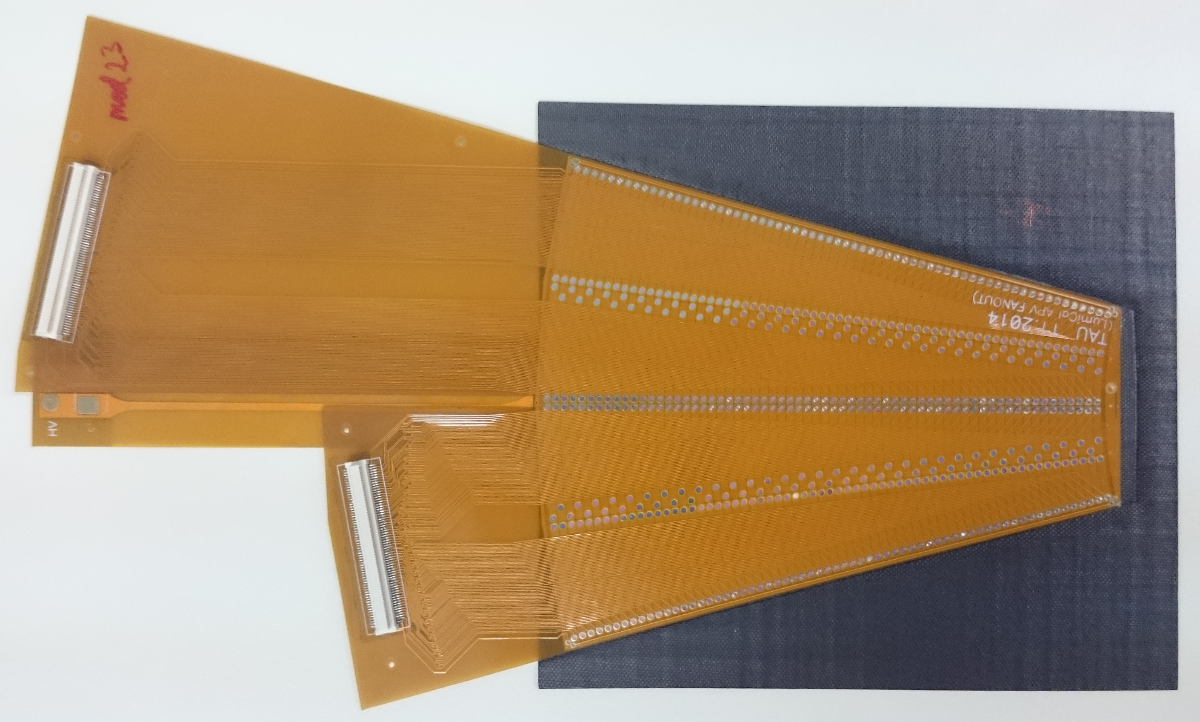}
    \caption{A thin LumiCal module. The black part is the carbon fiber support. The silicon sensor is covered by the Kapton-copper fan-out which has two connectors for front-end boards.}
    \label{module_photo}
  \end{minipage}\hfill
  \hspace{0.025\textwidth}
  \begin{minipage}[t]{0.47\textwidth}
    \includegraphics[width=\textwidth]{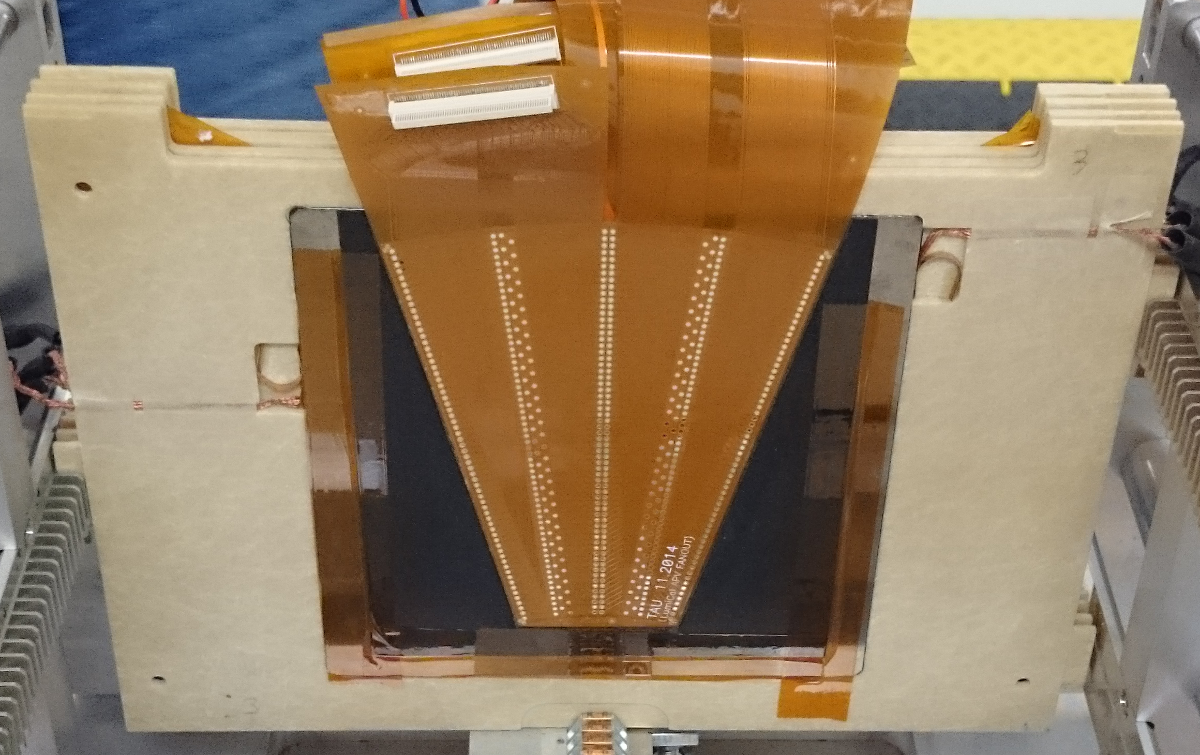}
    \caption{A thin LumiCal module attached by adhesive tape to the tungsten plate fixed in the permaglass frame.}
    \label{module_tungsten_photo}
  \end{minipage}
\end{figure}
The LumiCal prototype and tracker are assembled in a mechanical frame~\cite{mech_frame} specially designed to provide high precision of the positioning of the sensors and absorber planes and high flexibility to build a calorimeter prototype. A picture of the thin LumiCal module is shown in Fig.~\ref{module_photo}. It is attached to the tungsten absorber plate by adhesive tape (Fig.~\ref{module_tungsten_photo}). The tungsten plate is permanently glued to the permaglass frame and fixed in the comb slots of the mechanical structure. The assembly of the LumiCal prototype is illustrated in Fig.~\ref{tb2016_LumiCal}. Two thin LumiCal modules, viewed separately in the upper part of Fig.~\ref{tb2016_LumiCal}, are the tracker planes denoted as ``Tracker'' in Fig.~\ref{tb2016}. They are installed in front of the calorimeter. Two APV25 hybrid front-end boards are connected to each module as shown in Fig.~\ref{apv_connected_photo} when viewed from the side of short outputs of the fan-out. The stack is finally rotated so that the radial direction of the sensors with fine granularity is in the horizontal plane (perpendicular to the magnetic field) to provide more accurate measurements of electron deflection angle in the magnetic field.

About half a million events with each trigger configuration are collected. In addition, the LumiCal prototype response to the electron beam with energies from 1~GeV to 6~GeV, with 1~GeV step is recorded. The data analysis is in progress. 

\begin{figure}[h!]
  \includegraphics[width=\textwidth]{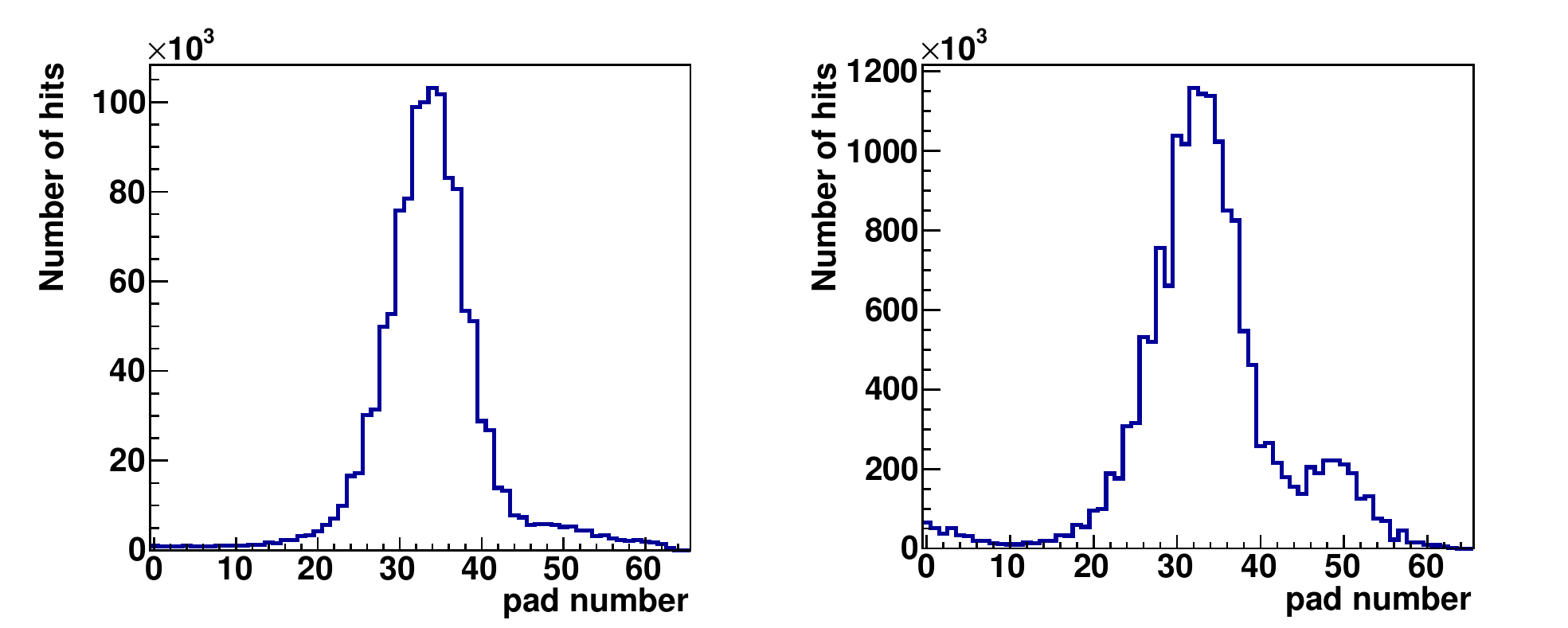}
  \caption{Occupancy of the tracker (left) and calorimeter (right) for the runs with a low energy photon trigger.}
  \label{occupancy}
\end{figure}

Figure~\ref{occupancy} shows preliminary results on the combined occupancy in the tracker (left) and in the calorimeter (right) for data collected with a low energy photon trigger. The histograms are filled with unity weight by the pad number of pads which observe the signal. It demonstrates the separation between electrons which produce a peak with a maximum around pad 34 both for the tracker and the calorimeter, and photons which produce a peak only in the calorimeter with a maximum around pad 50. Significant overlap between peaks in the calorimeter is explained by statistical fluctuations in the shower development and the transverse beam size, which can be taken into account after reconstructing positions of the showers in the calorimeter and electron position in telescopes T1 and T2. 
%
\section{Summary}
\label{Summary_section}
New submillimeter thickness detector modules for the luminosity calorimeter LumiCal have been designed and produced. Silicon sensors are readout using Kapton fan-outs with copper strips connected via wire bonding or TAB to the sensor pads. The assembled modules were successfully installed in the 1~mm gap between the tungsten absorbers during the 2015 and 2016 test-beam campaigns. The calorimeter prototypes with four and six sensitive layers were tested and the data analysis is in progress.

The recent development of the LumiCal detector demonstrates the feasibility of constructing a compact calorimeter consistent with the conceptual design, which is optimized for a high precision luminosity measurement in~e$^{+}$e$^{-}$ collider experiments. The experience and the techniques developed for the LumiCal module construction can be well used for other FCAL subsystems.
%
\section{Acknowledgments}
This work is partly supported by the Israel Science Foundation (ISF), the I-CORE program of the Israel Planning and Budgeting Committee. This project has received funding from the European Union's Horizon 2020 Research and Innovation programme under Grant Agreement no. 654168.

\end{document}